\documentclass[twocolumn,showpacs,amsmath,amssymb,prc,superscriptaddress,nofootinbib,floatfix]{revtex4}

\usepackage{graphicx}
\usepackage{amsfonts}
\usepackage{color}
\usepackage{txfonts}

\newcommand{\dd}{\mbox{$\textrm{d}$}}

\begin{document}
\renewcommand{\theenumi}{(\alph{enumi})}

\title{$\eta$-$^4{\rm He}$ interaction from the $d d \to \eta ^4{\rm He}$ reaction near threshold}

\date{\today}

\author{Ju-Jun Xie}~\email{xiejujun@impcas.ac.cn}%
\affiliation{Institute of Modern Physics, Chinese Academy of
Sciences, Lanzhou 730000, China}
\author{Wei-Hong Liang}~\email{liangwh@gxnu.edu.cn}%
\affiliation{Department of Physics, Guangxi Normal University,
Guilin 541004, China}
\author{Eulogio Oset}~\email{oset@ific.uv.es}
\affiliation{Departamento de F\'{\i}sica Te\'orica and IFIC, Centro
Mixto Universidad de Valencia-CSIC Institutos de Investigaci\'on de
Paterna, Aptdo. 22085, 46071 Valencia, Spain}%

\begin{abstract}

We analyze the data on the total cross sections for the $dd
\to \eta{}^4{\rm He}$ reaction close to threshold and look for
possible $\eta ^4 {\rm He}$ bound states. We develop a framework in
which the $\eta ^4{\rm He}$ optical potential is the key ingredient,
rather than parameterizing the scattering matrix, as is usually
done. The strength of this potential, together with some production
parameters, are fitted to the available experimental data. The
relationship of the scattering matrix to the optical potential is
established using the Bethe-Salpeter equation and the $\eta ^4{\rm
He}$ loop function incorporates the range of the interaction given
by the experimental $^4 {\rm He}$ density. However, when we look for poles of the scattering matrix, we get poles in the bound region, poles in
the positive energy region or no poles at all. If we further restrict the results with constraints from a theoretical model with all its uncertainties the bound states are not allowed. However, we find a bump structure
in $|T|^2$ of the $\eta ^4{\rm
He} \to \eta ^4{\rm
He}$ amplitude below threshold for the remaining solutions.

\end{abstract}


\maketitle

\section{Introduction}

The search for $\eta$ bound states in nuclei has been a constant
thought for several
years~\cite{Wilkin:2016ajz,Bass:2015ova,Haider:2015fea,Kelkar:2015bta,Hirenzaki:2015eoa,Friedman:2013zfa},
starting from the early works of
Refs.~\cite{Bhaleliu,Haiderliu,Liuhaider}. Follow up evaluations of
the $\eta$-nucleus optical potential, with special attention to
two-nucleon $\eta$ absorption, showed that, while indeed the
$\eta N$ interaction was strong enough to bind $\eta$ states, the
widths were always bigger than the binding~\cite{chiang}.

The $\eta$-nucleus interaction within the chiral unitary approach
was studied in Ref.~\cite{Inoueta}, where enough attraction was
found to form bound $\eta$-nucleus states. Detailed studies of the
$\eta$ energies for different nuclei were made in
Ref.~\cite{GarciaRecio:2002cu} where, for medium and light nuclei,
bound states were found (see also Ref.~\cite{Cieply:2013sga}, where
qualitatively similar conclusions were drawn), though with larger
widths than binding energies. For instance, some theoretical
calculations for light systems predicted binding energy $B_E$ of
around 1 MeV or less and width $\Gamma = 15$~MeV for $\eta ^3{\rm
He}$~\cite{Barnea:2015lia}. In the recent work~\cite{Barnea:2017oyk}
the results of Ref.~\cite{Barnea:2015lia} have been updated, in
particular the new width is much smaller. In
Ref.~\cite{Xie:2016zhs}, the data on cross sections and asymmetries
for the {\mbox{$pd \to \eta{}^3{\rm He}$}} reaction close to
threshold were studied with the aim of looking for bound states of the
$\eta ^3 {\rm He}$ system. The resulting $\eta ^3{\rm He}$
scattering matrix had a local Breit-Wigner form in a narrow range of
energies which corresponded to a binding of about $0.3$ MeV and a
width of about $3$ MeV. However, the pole appeared in the continuum,
not in the bound region.

The fact that the widths are expected to be much larger than the
binding might be the reason why so far, we have no conclusive
evidence for any of these bound
states~\cite{Bilger:2002aw,Mersmann,Colin,Rausmann:2009dn,Urban:2009zzc,
Chrien:1988gn,Johnson:1993zy,Frascaria:1994va,Willis:1997ix,Wronska:2005wk,Skurzok:2011aa,Adlarson:2013xg,Krzemien:2014ywa,Krzemien:2015fsa,Skurzok:2016fuv,Adlarson:2016dme,Sokol:1998ua,Budzanowski:2008fr,
Moskal:2010ee,Pheron:2012aj,Fujioka:2015pla}.

The first measurements of the $dd \to \eta ^4{\rm He}$ total cross
sections were carried out using the SPES4~\cite{Frascaria:1994va}
and the SPES3~\cite{Willis:1997ix} spectrometers at SATURNE. Later
on, the{$dd \to \eta{} ^4 {\rm He}$ reaction has been
investigated near threshold using the ANKE
facility~\cite{Wronska:2005wk}. The total cross sections have been
measured at two excess energies, $Q = 2.6$ and $7.7$ MeV. The data
on the $dd \to \eta{} ^4 {\rm He}$ total cross section show
a clear enhancement from threshold before becoming stable at an
excess energy of about $Q=3$~MeV, keeping this constant value up to
about 10~MeV~\cite{Willis:1997ix}. Recently, the search for $\eta
^4{\rm He}$ bound state of the $dd \to (\eta + ^4{\rm
He})_{\rm bound} \to X$ reaction has been proposed and performed
at the WASA-at-COSY
facility~\cite{Skurzok:2011aa,Adlarson:2013xg,Krzemien:2014ywa,Krzemien:2015fsa,Skurzok:2016fuv,Adlarson:2016dme}.
These measurements have been analyzed in Ref.~\cite{Ikeno:2017xyb}
within a theoretical model. The authors of Ref.~\cite{Ikeno:2017xyb}
used a phenomenological method with an optical potential for the
$\eta$-$^4{\rm He}$ interaction. The available data on the $dd \to
\eta ^4{\rm He}$ reaction are reproduced quite well for a broad range
of optical potential parameters, for some of which the authors
predicted the $\eta$-$^4{\rm He}$ bound state formation in the
subthreshold region. Furthermore, the theoretical calculations of
Ref.~\cite{Ikeno:2017xyb} were compared with the experimental data
below the $\eta$ production threshold, with the WASA-at-COSY
excitation functions for the {\mbox{$dd \to ^3 \!\! {\rm He}N\pi$}}
reactions in Ref.~\cite{Skurzok:2018paa}, where no clear signal of
the $\eta ^4{\rm He}$ bound state was found. As a consequence, the
analysis in Ref.~\cite{Skurzok:2018paa} made further strong
constraints on the $\eta$-$^4 {\rm He}$ optical potential. With the
results obtained in Ref.~\cite{Skurzok:2018paa}, most predictions of
an $\eta ^4 {\rm He}$ bound state seem to be excluded.

In the present work, we use an alternative method of analysis,
following the algorithms used in the chiral unitary approach. Our approach does not assume any particular form
of the amplitude, instead it is generated from an $\eta
^4{\rm He}$ potential which is fitted to the data. The $T$-matrix
then arises from the solution of the Lippmann-Schwinger equation,
although we use the Bethe-Salpeter equation (BSE) for convenience,
which allows us to keep relativistic terms, yet, ignoring only the
negative energy component of the, more massive, nucleon propagator.

As we
shall see later, the output of our calculations leads to an $\eta
^4{\rm He}$ optical potential. With this optical potential we solve the BSE for the
$\eta ^4\textrm{He}$ system, and look for poles of the $\eta ^4{\rm He} \to \eta ^4{\rm He}$ amplitude. We find that in some cases there are poles in the bound region, in other cases the poles are in the continuum and in other cases there are no poles.

Steps in a similar direction to ours were taken in
Ref.~\cite{Ikeno:2017xyb}, where the available data on the
$dd \to \eta{}^4{\rm He}$ reaction were studied in terms of
optical potentials. The results obtained here are similar to those
obtained in that work. Our study allows to see the statistical distributions of the values of the real and imaginary part of the fitted potential, and the correlation between them. The formalism is also different. In addition we show results when we put constraints from a theoretical model, yet allowing large uncertainties.

\section{Formalism}

In this section, we consider the $dd \to \eta ^4{\rm He}$ reaction
and explain our theoretical approach developed in the present work.
In Fig.~\ref{fig:fig1} we depict diagrammatically the $dd \to
\eta{}^4{\rm He}$ process.

\begin{figure*}[htbp]\centering
\hspace{-5.cm}
\includegraphics[scale=0.5]{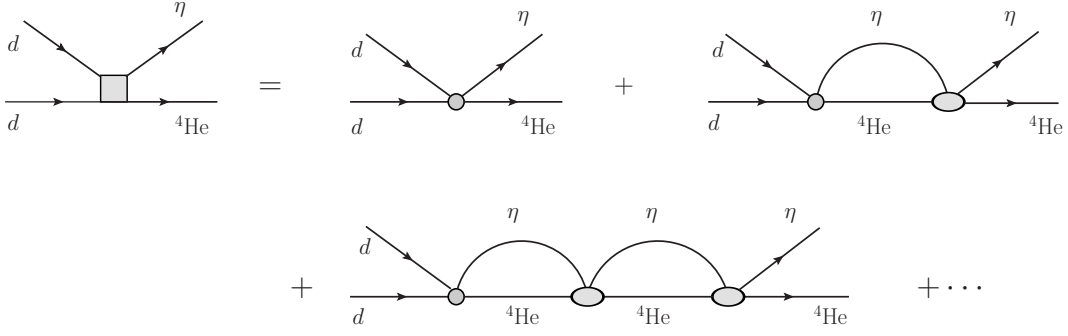}
\caption{The process $dd \to \eta{}^4{\rm He}$ considering
explicitly the $\eta ^4{\rm He}$ re-scattering. The square box in
the first diagram indicates the full transition amplitude, while the
circle in the second diagram stands for the bare transition
amplitude prior to the $\eta ^4{\rm He}$ final state interaction.
It contains all diagrams that do not have $\eta ^4{\rm He}$ as
intermediate state. The oval stands for the $\eta ^4{\rm He}$
optical potential. \label{fig:fig1}}
\end{figure*}

\subsection{The $\eta$-$^4\rm{He}$ interaction}

The $\eta ^4\textrm{He} \to \eta ^4\textrm{He}$ scattering amplitude
is given by the diagrams shown in Fig.~\ref{fig:fig2}, and formally
by the BSE
\begin{equation}\label{eq:BSE}
T= V+VGT,
\end{equation}
where $G$ is the loop function of intermediate $\eta ^4{\rm He}$ states, and $V$ is
the $\eta ^4\textrm{He}$ optical potential, which contains an
imaginary part to account for the inelastic channels $\eta
^4\textrm{He} \to X$ with $X$ being mostly $\pi ^3{\rm He} N$. It
also includes the $^3{\rm He}N^*(1535)$ intermediate state arising
mainly from the $\eta$ meson absorption, $\eta N \to
N^*(1535)$~\cite{Adlarson:2016dme}.

\begin{figure*}[htbp]
\begin{center}
\hspace{-6.cm}
\includegraphics[scale=0.5]{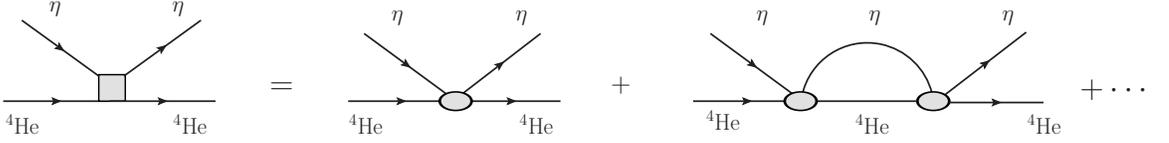}
\caption{Diagrammatic representation of the $\eta ^4\textrm{He} \to
\eta ^4{\rm He}$ scattering matrix. \label{fig:fig2}}
\end{center}
\end{figure*}

The low density theorem in many-body theory tells us that at low
densities the optical potential is given by
\begin{eqnarray}
\label{eq:vroptical} V(\vec{r}) = 4 t_{\eta N}
\tilde{\rho}(\vec{r}), \label{eq:potential}
\end{eqnarray}
where $t_{\eta N}$ is the forward $\eta N$ amplitude and
$\tilde{\rho}(\vec{r})$ is the $^4\rm{He}$ density normalized to
unity. Eq.~\eqref{eq:vroptical} is relatively accurate in many body
physics, but here we do not use it. We only take from it the density
dependence which provides a realistic range of the $\eta$-nucleus
interaction, since the $\eta$ can interact with all the nucleons in
the nucleus distributed according to $\rho(\vec{r})$.

In momentum space the potential is given by
\begin{eqnarray}
\nonumber V(\vec{p}_\eta,\vec{p'}_\eta) &=& 4t_{\eta N} \int
\dd^3\vec{r}\, \tilde{\rho}(\vec{r}) e^{i(\vec{p}_\eta -
\vec{p'}_{\eta})\cdot \vec{r}}\\ &=& 4t_{\eta N}F(\vec{p}_\eta -
\vec{p'}_{\eta}), \label{eq:opotential}
\end{eqnarray}
where $F(\vec{q})$ is the $^4\rm{He}$ form factor,
\begin{eqnarray}
F(\vec{q}) = \int \dd^3\vec{r}\, \tilde{\rho}(\vec{r}) e^{i\vec{q} \cdot
\vec{r}},
\end{eqnarray}
and $F(\vec{0}) = 1$. A good approximation to this form factor at small
momentum transfers is given by a Gaussian,
\begin{eqnarray}
F(\vec{q}) = e^{-\beta^2 |\vec{q}|^{2}},
\end{eqnarray}
where $\beta^2= \langle r^2\rangle/6$. This mean-square radius
corresponds to the distribution of the centers of the nucleons and,
after correcting for the nucleon size, it leads to an experimental
value of $\beta^2 = 12.1 ~{\rm GeV}^{-2}$ which was obtained with
$\langle r^2\rangle^{1/2}_{^4{\rm He}} = 1.68$ fm as in
Ref.~\cite{Sick:2014yha}.

Because of this form factor, the optical potential in
Eq.~\eqref{eq:opotential} contains all partial waves. After
integrating over  the angle between $\vec{p}^{~'}_\eta$ and
$\vec{p}_\eta$, the $s$-wave projection of the optical potential
becomes
\begin{eqnarray}
V(\vec{p}_\eta,\vec{p}^{~'}_\eta) & \!\!= \!\! & 4 t_{\eta N}
\frac{1}{2}\int^1_{-1} \!\!\! \dd\cos\theta e^{-\beta^2
(|\vec{p}_\eta|^2 + |\vec{p'}_\eta|^2 - 2|\vec{p}_{\eta}|
|\vec{p'}_{\eta}|\cos\theta)} \nonumber\\
&&\hspace{-1.7cm} = 4 t_{\eta N} e^{-\beta^2 |\vec{p}_{\eta}|^2}
e^{-\beta^2 |\vec{p'}_{\eta}|^2} \left[1 + \frac{1}{6}
(2\beta^2|\vec{p}_{\eta}||\vec{p'}_{\eta}|)^2 + ... \right]\!.
\end{eqnarray}
One can easily see that the terms
$(2\beta^2|\vec{p}_{\eta}||\vec{p'}_{\eta}|)^2/6 + ...$ are
negligible in the region where $e^{-\beta^2 |\vec{p}_{\eta}|^2}
e^{-\beta^2 |\vec{p'}_{\eta}|^2}$ is sizeable and this leads to a
potential that is separable in the variables $\vec{p}_\eta$ and
$\vec{p}^{~'}_\eta$, which makes the solution of Eq.~\eqref{eq:BSE}
trivial. Keeping the relativistic factors of the BSE, we can
write~\cite{Oset:1997it}:
\begin{eqnarray}
T(\vec{p}_\eta,\vec{p'}_\eta) &=& \tilde{V} e^{-\beta^2
|\vec{p}_{\eta}|^2} e^{-\beta^2 |\vec{p'}_{\eta}|^2}  + \\
\nonumber && \hspace{-2.2cm}\int \frac{\dd^3\vec{q}}{(2\pi)^3}
\frac{M_{^4{\rm He}}}{2\omega_{\eta}(\vec{q}) E_{^4{\rm
He}}(\vec{q})} \frac{\tilde{V} e^{-\beta^2 |\vec{p}_{\eta}|^2}
e^{-\beta^2 |\vec{q}\hspace{0.5mm}|^2}}{\sqrt{s} -
\omega_{\eta}(\vec{q}) - E_{^4{\rm He}}(\vec{q}) + i\epsilon}
T(\vec{q},\vec{p'}_\eta), \label{eq:Tamplitude}
\end{eqnarray}
with $\sqrt{s}$ being the invariant mass of the $\eta ^4{\rm He}$
system, $\omega_{\eta}(\vec{q}) = \sqrt{m^2_{\eta} +
|\vec{q}\hspace{0.5mm}|^2}$, and $E_{^4{\rm He}} (\vec{q}) =
\sqrt{M^2_{^4{\rm He}} + |\vec{q}\hspace{0.5mm}|^2}$. Note that here we
have taken $\tilde{V}$ instead of $4t_{\eta N}$ for more
generality.

The $T$ matrix can be factorized in the same way as $V$, and we
have~\cite{Xie:2016zhs}
\begin{eqnarray}
T(\vec{p}_\eta,\vec{p'}_\eta) = t e^{-\beta^2 |\vec{p}_{\eta}|^2}
e^{-\beta^2 |\vec{p'}_{\eta}|^2}. \label{eq:Tfull}
\end{eqnarray}
The BSE becomes then algebraic
\begin{eqnarray}
t = \tilde{V} + \tilde{V} G t, \label{eq:ttilde}
\end{eqnarray}
with a loop function
\begin{eqnarray}
&& G (\sqrt{s}) = \frac{M_{^4{\rm
He}}}{16\pi^3} \times \nonumber \\
&& \int \frac{\dd^3\vec{q}}{\omega_\eta(\vec{q})E_{^4{\rm
He}}(\vec{q})} \frac{e^{-2 \beta^2 |\vec{q}|^2}}{\sqrt{s} -
\omega_{\eta}(\vec{q}) - E_{^4{\rm He}}(\vec{q}) + i\epsilon}.
\label{eq:gfunction}
\end{eqnarray}

In Fig.~\ref{fig:gfunction}, we show the real and imaginary parts of
the loop function $G$ as a function of the excess energy $Q$ ($Q =
\sqrt{s} - m_\eta - M_{^4{\rm He}}$) with $M_{^4{\rm He}} = 3728.4$
MeV and $m_{\eta} = 547.862$ MeV. We can see a strong cusp of the
real part at the $\eta ^4{\rm He}$ threshold and the imaginary part
starting from this threshold.

\begin{figure}[htbp]\centering
\includegraphics[width=1.0\columnwidth]{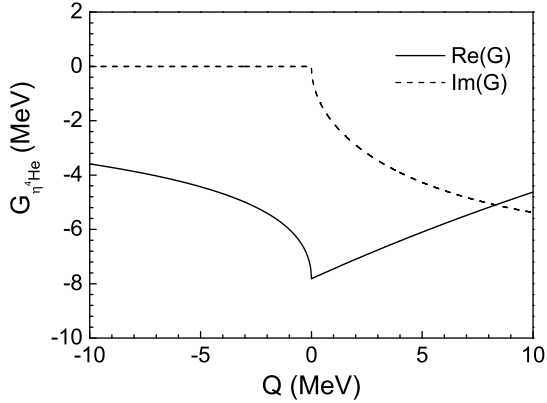}
\caption{Real (solid line) and imaginary (dashed line) parts of the $G$ of
Eq.~\eqref{eq:gfunction} as functions of the excess energy $Q$.
\label{fig:gfunction}}
\end{figure}

In the normalization that we are using, the $\eta$-nucleon and
$\eta$-$^4$He scattering lengths are related to the $t$-matrices by
\begin{eqnarray}
a_{\eta N} &=& \left.\frac{1}{4\pi}\frac{m_N}{\sqrt{s_{\eta N}}} t_{\eta N}
\right|_{\sqrt{s_{\eta N}} = m_N + m_\eta}\\
a_{\eta ^4{\rm He}} &=& \left.\frac{1}{4\pi}\frac{M_{^4{\rm
He}}}{\sqrt{s}} T \right|_{\sqrt{s} = M_{^4{\rm He}} + m_\eta}.
\label{eq:aetahe3}
\end{eqnarray}

The strategy that we adopt is to fit $\tilde{V}$ to the {\mbox{$dd
\to \eta{}^4{\rm He}$}} data and then see how different $\tilde{V}$
is from $4t_{\eta N}$ by evaluating
\begin{eqnarray}
a'_{\eta N} = \left.\frac{1}{4\pi}\frac{m_N}{\sqrt{s_{\eta N}}}
\frac{\tilde{V}}{4} \right|_{\sqrt{s_{\eta N}} = m_N + m_\eta}
\label{effective}
\end{eqnarray}
and comparing it to the theoretical value of $a_{\eta N}$.

After obtaining the \emph{best} value for $\tilde{V}$, we then plot
\begin{eqnarray}
T = t e^{-2\beta^2 |\vec{p}_{\eta}|^2}
\end{eqnarray}
and investigate it below threshold.

\subsection{Production amplitude}

Following the formalism of
Refs.~\cite{Wronska:2005wk,Ikeno:2017xyb}, we write for the
{\mbox{$dd \to \eta{}^4{\rm He}$}} transition depicted as a circle
in Fig.~\ref{fig:fig1}
\begin{eqnarray}
V_P = A (\vec{\epsilon}_1 \times \vec{\epsilon}_2 ) \cdot \vec{p}_d,
\label{eq:vp}
\end{eqnarray}
where $\vec{\epsilon}_1$ and $\vec{\epsilon}_2$ are the
polarizations of the initial two deuterons, and $\vec{p}_d$ is the
momentum in the initial state. This amplitude has the initial-state
$p$-wave needed to match the $\eta (0^-) ^4{\rm He} (0^+)$ with the
$d (1^+) d (1^+)$ system. This vertex accounts for all mechanisms of $dd \to \eta ^4{\rm He}$ reaction which do not have $\eta ^4{\rm He}$ as an intermediate step, as direct $dd \to \eta ^4{\rm He}$, $dd \to \pi ^3{\rm He} N \to \eta ^4{\rm He}$, $dd \to \pi NNNN \to \eta ^4{\rm He}$, \textit{etc}.

With similar arguments to those used to derive Eq.~\eqref{eq:opotential}, we
can justify that $V_P$ in Eq.~\eqref{eq:vp} must be accompanied by the factor
$e^{-\beta^2|\vec{p}_\eta|^2}$, which, if the $\eta$ is in the loop, will
become $e^{-\beta^2|\vec{q}|^2}$. In view of this we can write analytically
the equation for the diagrams of Fig.~\ref{fig:fig1} as,
\begin{eqnarray}
t_{dd \to \eta ^4{\rm He}} &=& V_P e^{-\beta^2 |\vec{p}_{\eta}|^2} +
V_P G t e^{-\beta^2 |\vec{p}_{\eta}|^2} \nonumber\\
& \! = \! & V_P e^{-\beta^2 |\vec{p}_{\eta}|^2} (1 + G t) \! = \!
\frac{V_P e^{-\beta^2 |\vec{p}_{\eta}|^2}}{1-\tilde{V}G},
\label{eq:tdptoetahe3}
\end{eqnarray}
where in the last step we have used Eq.~\eqref{eq:BSE}. The cross section
then becomes
\begin{eqnarray}
\sigma = \frac{M^2_d M_{^4{\rm He}}}{9 \pi s}
\frac{|A|^2}{|1-\tilde{V}G|^2} |\vec{p}_\eta| |\vec{p}_d|
e^{-2\beta^2 |\vec{p}_{\eta}|^2}, \label{eq:tcs-swava}
\end{eqnarray}
where we have used
\begin{eqnarray}
\label{eq:19} \overline{\sum} \sum |V_P|^2 = \frac{2}{9} |A|^2
|\vec{p}_d|^2.
\end{eqnarray}
This allows us to perform a fit to the data up to an excess energy
$Q=8.5$~MeV, and thus determine $\tilde{V}$. From this we shall
determine $T$ by means of Eqs.~\eqref{eq:Tfull} and
\eqref{eq:ttilde}, and investigate its structure below threshold.

We should note that we are taking both $A$ and $\tilde{V}$ constant. Certainly these magnitudes are energy dependent. The relevant data for the study of the $\eta ^4{\rm He} \to \eta ^4{\rm He}$ scattering amplitude are in a range of $2.5$ MeV where changes in these magnitudes should be negligible. The most critical case would be the magnitude $\tilde{V}$, related to $t_{\eta N}$, which is dominated by the $N^*(1535)$ pole. To estimate the changes in this magnitude we have used the model of Ref.~\cite{chiang} and looked at the dependence on the energy. We find that the changes in this magnitude in a range of $2.5$ MeV are of the order of $1-2\%$ playing with the uncertainties of the model. This uncertainty is much smaller than the errors that we find from a fit to the data and justify taking $\tilde{V}$ as a constant, and $A$ a fortiori since it is not driven by a resonance.

\section{Results}
\label{results}

We perform a three-parameter [$|A| = r_A$ and $\tilde{V} = {\rm
Re}(V) + i {\rm Im}(V)$] $\chi^2$ fit to the experimental data on
the total cross sections of the {\mbox{$dd \to \eta{} ^4{\rm He}$}}
reaction below $Q = 8.5$~MeV. There are 12 experimental data points
in total. The values of the resulting parameters are collected in
Table~\ref{tab:fittedparas}. One can see that the fitted parameters
have large uncertainties, especially for the real part of
$\tilde{V}$.

\begin{table}[htbp]
\caption{Values of parameters determined in this
work.}\label{tab:fittedparas}
\begin{tabular}{|c|c|}
\hline
Parameter   & Fitted value \\
\hline
 $r_A [{\rm MeV}^{-5/2}]$   & $(7.6 \pm 2.3) \times 10^{-9}$ \\
 ${\rm Re}(V) [{\rm MeV}^{-1}]$      & $(-12.3 \pm 18.4)\times 10^{-2}$    \\
 ${\rm Im}(V) [{\rm MeV}^{-1}]$       & $(-13.7 \pm 5.4)\times 10^{-2}$    \\
\hline
\end{tabular}
\end{table}

To get more precise information from the experimental measurements,
we generate random sets of the experimental data within the range of
error of each datum with a Gaussian distribution. For each set of data, we perform a $\chi^2$
fit, and the corresponding fitted parameters are determined by the
best fit. In this way, we get sets of the fitted parameters ($r_A$,
${\rm Re}(V)$, ${\rm Im}(V)$) with different best $\chi^2_{\rm
best}$~\cite{Landay:2016cjw,Perez:2014jsa}. With these best fits we get the
shaded region~\footnote{We remove $16\%$ in both extreme of the output to get $68\%$ confidence level.} shown in
Fig.~\ref{fig:tcsfull}, from where one can see that the experimental
data can be well reproduced. Besides, in Fig.~\ref{fig:tcsfull} we
also show the fitted total cross sections with the centroid values
of the fitted parameters listed in Table~\ref{tab:fittedparas} by
the red-solid curve. To get values for the observables we evaluate them
with the parameters of each fit and from there we get the average value
and the dispersion. In this way we take into account the correlations
between the parameters.

\begin{figure}[htbp]\centering
\includegraphics[scale=0.6]{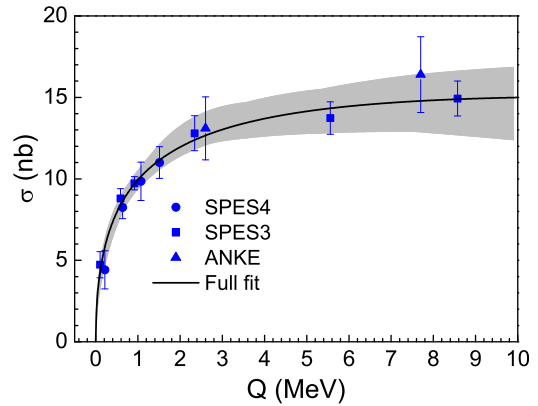}
\vspace{-6.5cm}
\caption{The fitted $dd \to \eta ^4{\rm He}$ total cross sections
compared with experimental data. The circles are taken from
Ref.~\cite{Frascaria:1994va}; squares are from
Ref.~\cite{Willis:1997ix}; and triangles are from
Ref.~\cite{Wronska:2005wk}. \label{fig:tcsfull}}
\end{figure}

On the other hand, as is general in particle physics, and in
particular in the case of the $\eta ^4{\rm He}$ scattering matrix,
the position of the poles of $T$ does not coincide with that of the
mass and width of a possible Breit-Wigner parametrization. We
investigate the position of the poles here. In table~\ref{tab:poles}
we show the position of the poles ($z_R = E - i\Gamma/2$) with the
energy $E$ measured from the $\eta ^4 {\rm He}$ threshold.

\begin{table}[htbp]
\caption{Pole position of the $T$ matrix for the $\eta ^4{\rm He}
\to \eta ^4{\rm He}$ scattering.}\label{tab:poles}
\begin{tabular}{|c|c|c|c|}
\hline
${\rm Re}(V) ~ [{\rm MeV}^{-1}]$   & ${\rm Im}(V) ~ [{\rm MeV}^{-1}]$ & $E$ [MeV] & $\Gamma$ [MeV] \\
\hline
 $-0.123$   & $-0.137$  & $10.0$  & $13.3$    \\
 $-0.2$      & $-0.15$  & $2.0$   & $26.9$    \\
 $-0.2$      & $-0.1$   & $-0.4$  & $16.2$    \\
 $-0.22$     & $-0.1$   & $-2.4$  & $17.8$    \\
 $-0.24$     & $-0.1$   & $-4.5$  & $19.2$    \\
 $-0.26$     & $-0.1$   & $-6.6$  & $20.4$    \\
 $-0.26$     & $-0.08$  & $-7.2$  & $16.1$    \\
 $-0.26$     & $-0.06$  & $-7.6$  & $11.9$    \\
 $-0.26$     & $-0.04$  & $-7.9$  & $7.9$     \\
 $-0.28$     & $-0.1$   & $-8.9$  & $21.4$    \\
\hline
\end{tabular}
\end{table}

We can see that for the centroid values of $\tilde{V}$ in
table~\ref{tab:fittedparas} we get a pole in the unbound region
around $10.6$ MeV and with $\Gamma$ around $11.4$ MeV. Stretching
the errors in ${\rm Re}(V)$ in table~\ref{tab:fittedparas}, if we
take ${\rm Re}(V) = -0.2$ MeV$^{-1}$ and ${\rm Im}(V) = -0.1 {\rm
MeV}^{-1}$ we find now a pole with $E$ around $-0.4$ MeV and width
$\Gamma$ of about $16.2$ MeV. To complete the table we show what
happens if we increase ${\rm Re}(V)$ in size in the range of
table~\ref{tab:fittedparas}. For ${\rm Re}(V) = -0.24 {\rm
MeV}^{-1}$ and ${\rm Im}(V) = -0.1 {\rm MeV}^{-1}$ we find already a
bound state around $E=-4.5$ MeV and with a width $\Gamma$ of about
$19.2$ MeV. In conclusion, because of the limited experimental data,
we cannot always find a bound state with the fitted potential, which
coincides with the conclusion of Ref.~\cite{Xie:2016zhs} for $\eta
^3{\rm He}$ and the conclusions of
Refs.~\cite{Ikeno:2017xyb,Skurzok:2018paa}.

We also observe the general trend that the widths are bigger than
the binding.

In order to understand the results of our analysis more clearly, we
plot three areas inside the ${\rm Re}(V)$-${\rm Im}(V)$ plane as
shown in Fig.~\ref{fig:potential}, where the acceptable region of
${\rm Re}(V)$ and ${\rm Im}(V)$ values can be easily understood for
having an $\eta ^4{\rm He}$ bound state. The poles in areas I and II
are in the unbound and bound regions, respectively, while there is
no pole in area III. In Fig.~\ref{fig:potential}, we show also the fitted results of the
pairs of ${\rm Re}(V)$ and ${\rm Im}(V)$, which correspond to those
best fits as discussed above.

\begin{figure}[htbp]\centering
\includegraphics[scale=0.6]{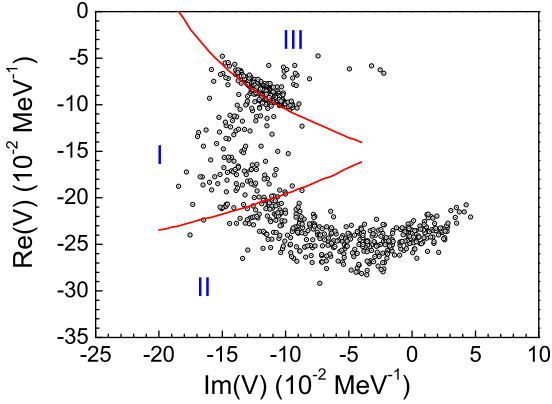}
\vspace{-6.5cm}
\caption{Poles area in the ${\rm Re}(V)$-${\rm Im}(V)$ plane, where
the circles are the best fitted pairs of [${\rm Re}(V), {\rm
Im}(V)$]. \label{fig:potential}}
\end{figure}

On the other hand, with the potentials $\tilde{V}$ obtained from the
best fits as shown in Fig.~\ref{fig:potential}, we have evaluated
the scattering length $a'_{\eta N}$ of Eq.~\eqref{effective}:
\begin{eqnarray}
a'_{\eta N} = [-(0.39 \pm 0.19) - i(0.23 \pm 0.12) ] ~~ {\rm fm},
\label{eq:aetaNresults}
\end{eqnarray}
which is comparable with the value obtained in
Ref.~\cite{Xie:2016zhs} within errors. The errors quoted here are
statistical and they are determined as the standard deviation.

Similarly, by means of Eq.~\eqref{eq:aetahe3}, we calculate the
$\eta ^4{\rm He}$ scattering length to be
\begin{eqnarray}
a_{\eta ^4{\rm He}} = [(2.11 \pm 1.07) - i(1.21 \pm 0.67) ] ~~ {\rm fm}.
\label{eq:aetahe3results}
\end{eqnarray}
The value of $a_{\eta ^4{\rm He}}$ obtained here is different from
the results obtained in
Refs.~\cite{Willis:1997ix,Wronska:2005wk,Ikeno:2017xyb}, while the
absolute value of $a_{\eta ^4{\rm He}}$ is compatible with the
results of these works.

Note that the strategy of fitting an optical potential to the data
instead of the usual $t$-matrix parametrization used in previous
works, allows us to determine the sign of the real part of the
scattering lengths.

The fit done here produces an attractive potential, which is
consistent with all theoretical derivations of $t_{\eta N}$,
together with the $t_{\eta N} \tilde{\rho}(\vec{r})$ assumption for
the optical potential.

One should also note that in Ref.~\cite{Xie:2016zhs} one not only
fitted the $pd \to \eta ^3{\rm He}$ total cross sections but also
the asymmetry parameter in terms of the $\eta$ momentum. Also the
quality of the data of $pd \to \eta ^3{\rm He}$ reaction is much
better than for the present reaction. As a consequence, we could
determine the parameters in the case of the $\eta ^3{\rm He}$
interaction with more precision than in the present case.

Since we have less precision than in the $p d \to \eta ^3{\rm He}$ reaction we investigate what happens by adding more theoretical constraints. For this we assume now that Eq.~\eqref{eq:potential} is changed to
\begin{eqnarray}
V(r') = 4 \bar{t}_{\eta N} \tilde{\rho}(r'),
\end{eqnarray}
where $\bar{t}_{\eta N}$ is the $\eta N \to \eta N$ scattering amplitude modified in the medium, obtained in Ref.~\cite{chiang}, assuming that the driving term for the $\bar{t}_{\eta N}$ amplitude is given by the excitations of the $N^*(1535)$ and changes in the mass of $N^*(1535)$ and its width are calculated within many body theory. Then we make random choices of all the variables in the model within their uncertainties. The $\bar{t}_{\eta N}$ amplitude is now given by~\cite{chiang}
\begin{eqnarray}
\bar{t}_{\eta N} &=& \frac{g^2_{\eta}}{\sqrt{s_{\eta N}} - M_{N^*} + i\Gamma_{N^*}/2 - \delta_{M^*} + i \delta_{\Gamma_{N^*}}/2}, \label{eq:tbaretaN}
\end{eqnarray}
with
\begin{eqnarray}
  g^2_{\eta} &=& \frac{2\pi \Gamma_{N^*} {\rm Br}(N^* \to \eta N) M_{N^*}}{m_N q_\eta}, \\
  q_\eta &=& \frac{\sqrt{[M^2_{N^*} - (m_N+m_\eta)^2][M^2_{N^*}-(m_N - m_\eta)^2]}}{2M_{N^*}},
\end{eqnarray}
and we take
\begin{eqnarray}
M_{N^*} &=& 1530 \pm 15 ~~{\rm MeV}, \\
  \Gamma_{N^*} &=& 150 \pm 25 ~~{\rm MeV}, \\
 \delta_{M_{N^*}} &=& \pm 50 ~~ {\rm MeV}, \\
 \delta_{\Gamma_{N^*}} &=& (0-70) ~~ {\rm MeV}, \\
   {\rm Br}(N^* \to \eta N) &=& (42.5 \pm 12.5)\% .
\end{eqnarray}

The results of ${\rm Re(V)} \equiv {\rm Re}(4\bar{t}_{\eta N})$ and ${\rm Im(V)} \equiv {\rm Im}(4\bar{t}_{\eta N})$ are shown in Fig.~\ref{fig:potential-constraints}. We see that we get a fair distribution of possible pairs of (${\rm Re}(V)$, ${\rm Im}(V)$) playing with the uncertainties. Yet, the overlap with the distribution of Fig.~\ref{fig:potential}, shows that only part of the solutions with poles in the continuum and the solutions with no poles are acceptable. The bound solutions are far away from the overlap region.

\begin{figure}[htbp]\centering
\includegraphics[scale=0.6]{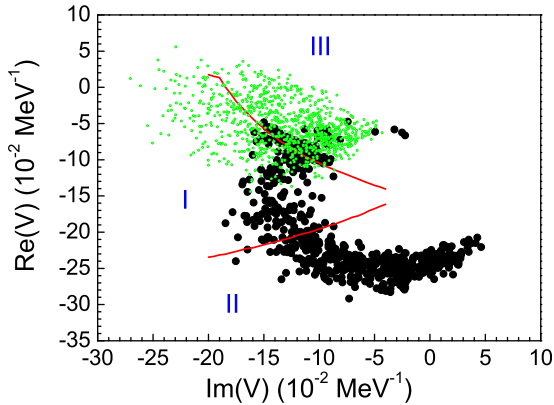}
\vspace{-6.5cm}
\caption{(color online) Poles area in the ${\rm Re}(V)$-${\rm Im}(V)$ plane, where
the blacks dots are the best fitted pairs of [${\rm Re}(V), {\rm
Im}(V)$] and the green circles are obtained with Eq.~\eqref{eq:tbaretaN}. \label{fig:potential-constraints}}
\end{figure}

\begin{figure}[htbp]\centering
\includegraphics[scale=0.6]{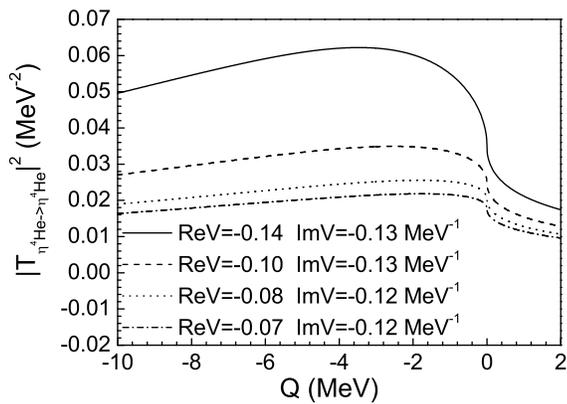}
\vspace{-6.5cm}
\caption{Square of the absolute value of the $\eta ^4{\rm He} \to
\eta ^4{\rm He}$ scattering amplitude. \label{fig:tsquare}}
\end{figure}

We have taken a sample of potentials in the overlap region of Fig.~\ref{fig:potential-constraints} and evaluate $T$ and $|T|^2$. The numerical results for the $|T|^2$ are shown in Fig.~\ref{fig:tsquare}. We see that in all cases there is a bump structure below the $\eta ^4{\rm He}$ threshold, which is tied
to the fast increase of the $dd \to \eta ^4{\rm He}$ cross sections
close to the reaction threshold.

In Fig.~\ref{fig:tetahe4}, we show separately the real and
imaginary parts of $T$ for these solutions.
We see that around $Q = -8$~MeV, ${\rm Re}(T)$ passes through zero, and $|{\rm
Im}(T)|$ has a bump around $-4$ MeV. This structure is reminding of a Breit-Wigner form, but somewhat distorted since the position of the maximum of $|{\rm Im}(T)|$ and the zero of ${\rm Re}(T)$ do not coincide.

\begin{figure}[htbp]
\centering
\includegraphics[scale=0.6]{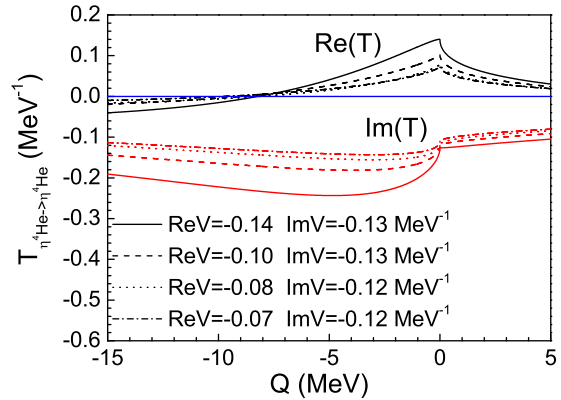}
\vspace{-6.5cm}
\caption{(color online) Real and imaginary parts of the $\eta ^4{\rm
He} \to \eta ^4{\rm He}$ amplitude $T$ as a function of the excess
energy $Q$. \label{fig:tetahe4}}
\end{figure}

\section{Summary and conclusions}

We have performed an analysis of the data on the total cross sections of
the {\mbox{$dd \to \eta{}^4 {\rm He}$}} reaction close to threshold.
Unlike former approaches that make a parametrization of the
amplitude, we express the total cross sections in terms of an
optical potential from which the $\eta ^4 {\rm He}$ scattering
amplitude is evaluated. The $T$ matrix is evaluated from the
potential using the Bethe-Salpeter equation and the loop function
$G$ of the intermediate $\eta ^4 {\rm He}$ state. This reflects the
range of the $\eta ^4 {\rm He}$ interaction, as given by the
empirical density of the $^4 {\rm He}$ nucleus.

The potential and other parameters related to the production
vertices are fitted to the data and with this potential we search for possible poles of the $\eta ^4{\rm He}$ system. We found poles in the unbound energy region, in the bound region or
no poles at all. We also obtain an $\eta ^4 {\rm
He}$ scattering length of the order of $[(2.11 \pm 1.07) - i(1.21 \pm
0.67) ] ~ {\rm fm}$.

In order to put restrictions on the solutions obtained from the $dd \to \eta ^4{\rm He}$ total cross sections we used a theoretical model for the $\eta ^4{\rm He}$ scattering amplitude in the nuclear medium based on the excitation of the $N^*(1535)$ by $\eta N$ and the medium modifications of this resonance studied in Ref.~\cite{chiang}, playing with all uncertainties in the parameters of the model. In this way we obtained a relatively wide region of values for the $\eta ^4{\rm He}$ potential, and the overlap with the solutions from the analysis of the data on the $dd \to \eta ^4{\rm He}$ reaction eliminated many of the solutions allowed by $dd \to \eta ^4{\rm He}$ alone. Taking the solutions from the overlap region, we could see that in these cases there was a bump structure of $|T|^2$ below threshold, closely related to the
shape of the $\eta ^4 {\rm He}$ production cross sections close to
threshold.

In summary, the new approach to the analysis of the {\mbox{$dd \to
\eta{}^4{\rm He}$}} data close to threshold has proved quite useful
and has been able to provide information on $\eta ^4 {\rm He}$ interaction.

It remains to be seen if the structure found below threshold could be seen in some experiments. The broad shape of $|T|^2$ in
Fig.~\ref{fig:tsquare} would not make
the matter easy, and in addition one should take into account that
large contributions of background source from reactions not tied to
the direct $\eta ^4{\rm He}$ interaction would further blur a possible signal. A message we found from the analysis of $dd \to \eta ^4{\rm He}$ data is that due to the limited
data and the large errors in the fit, we find that these data cannot
confirm nor rule out the existence of poles in the bound region,
which would correspond to $\eta ^4{\rm He}$ bound states. In any
case, when bound states appear we still see the general
rule that the widths are larger than the binding energies.

The last part of our investigation was to combine the $dd \to \eta ^4{\rm He}$  data with a theoretical model accommodating large uncertainties and from the overlap of the potentials allowed by this model and the data of $dd \to \eta ^4{\rm He}$ reaction we could see that the solutions leading to bound states were rejected. Yet, for the remaining solutions there was always a broad structure of $|T|^2$ below threshold independently
that the potentials lead to poles in the continuum or no poles at all.

\section*{Acknowledgments}

We would like to express our thanks to Michael D\"{o}ring for useful
discussions with him. This work is partly supported by the National
Natural Science Foundation of China (Grants No.\ 11475227, No.\
11735003, No.\ 11565007, and No.\ 11747307) and the Youth Innovation
Promotion Association CAS (No.\ 2016367). This work is also partly
supported by the Spanish Ministerio de Economia y Competitividad and
European FEDER funds under the contract number FIS2014-57026-REDT,
FIS2014-51948-C2- 1-P, and FIS2014-51948-C2-2-P, and the Generalitat
Valenciana in the program Prometeo II-2014/068.

\bibliographystyle{plain}

\begin{thebibliography}{999}

\bibitem{Wilkin:2016ajz}
  C.~Wilkin,
  Acta Phys.\ Polon.\ B {\bf 47}, 249 (2016).

\bibitem{Bass:2015ova}
  S.~D.~Bass and P.~Moskal,
  Acta Phys.\ Polon.\ B {\bf 47}, 373 (2016).

\bibitem{Haider:2015fea}
  Q.~Haider and L.~C.~Liu,
  Int.\ J.\ Mod.\ Phys.\ E {\bf 24}, 1530009 (2015).

\bibitem{Kelkar:2015bta}
  N.~G.~Kelkar,
  Acta Phys.\ Polon.\ B {\bf 46}, 113 (2015).

\bibitem{Hirenzaki:2015eoa}
  S.~Hirenzaki, H.~Nagahiro, N.~Ikeno, and J.~Yamagata-Sekihara,
  Acta Phys.\ Polon.\ B {\bf 46}, 121 (2015).

\bibitem{Friedman:2013zfa}
  E.~Friedman, A.~Gal, and J.~Mare${\rm \check{s}}$,
  Phys.\ Lett.\ B {\bf 725}, 334 (2013).

\bibitem{Bhaleliu}
  R.~S.~Bhalerao and L.~C.~Liu,
  Phys.\ Rev.\ Lett.\  {\bf 54}, 865 (1985).

\bibitem{Haiderliu}
  Q.~Haider and L.~C.~Liu,
  Phys.\ Lett.\ B {\bf 172}, 257 (1986).

\bibitem{Liuhaider}
  L.~C.~Liu and Q.~Haider,
  Phys.\ Rev.\ C {\bf 34}, 1845 (1986).

\bibitem{chiang}
  H.~C.~Chiang, E.~Oset, and L.~C.~Liu,
  Phys.\ Rev.\ C {\bf 44}, 738 (1991).

\bibitem{Inoueta}
  T.~Inoue and E.~Oset,
  Nucl.\ Phys.\ A {\bf 710}, 354 (2002).


\bibitem{GarciaRecio:2002cu}
  C.~Garcia-Recio, J.~Nieves, T.~Inoue, and E.~Oset,
  Phys.\ Lett.\ B {\bf 550}, 47 (2002).

\bibitem{Cieply:2013sga}
  A.~Ciepl\'{y}, E.~Friedman, A.~Gal, and J.~Mare\v{s},
  Nucl.\ Phys.\ A {\bf 925}, 126 (2014).


\bibitem{Barnea:2015lia}
  N.~Barnea, E.~Friedman, and A.~Gal,
  Phys.\ Lett.\ B {\bf 747}, 345 (2015).

\bibitem{Barnea:2017oyk}
  N.~Barnea, E.~Friedman and A.~Gal,
  Nucl.\ Phys.\ A {\bf 968}, 35 (2017).

\bibitem{Xie:2016zhs}
  J.~J.~Xie, W.~H.~Liang, E.~Oset, P.~Moskal, M.~Skurzok and C.~Wilkin,
  Phys.\ Rev.\ C {\bf 95}, 015202 (2017).


\bibitem{Bilger:2002aw}
  R.~Bilger {\it et al.},
  Phys.\ Rev.\ C {\bf 65}, 044608 (2002).

\bibitem{Mersmann}
  T.~Mersmann {\it et al.},
  Phys.\ Rev.\ Lett.\  {\bf 98}, 242301 (2007).

\bibitem{Colin}
  C.~Wilkin {\it et al.},
  Phys.\ Lett.\ B {\bf 654}, 92 (2007).

\bibitem{Urban:2009zzc}
  J.~Urban {\it et al.} [COSY-GEM Collaboration],
  Int.\ J.\ Mod.\ Phys.\ A {\bf 24}, 206 (2009).

\bibitem{Frascaria:1994va}
  R.~Frascaria {\it et al.},
  Phys.\ Rev.\ C {\bf 50}, R537 (1994).

\bibitem{Willis:1997ix}
  N.~Willis {\it et al.},
  Phys.\ Lett.\ B {\bf 406}, 14 (1997).

\bibitem{Wronska:2005wk}
  A.~Wronska {\it et al.},
  Eur.\ Phys.\ J.\ A {\bf 26}, 421 (2005).
\bibitem{Skurzok:2011aa}
  M.~Skurzok, P.~Moskal and W.~Krzemien,
  Prog.\ Part.\ Nucl.\ Phys.\  {\bf 67}, 445 (2012).
\bibitem{Adlarson:2013xg}
  P.~Adlarson {\it et al.} [WASA-at-COSY Collaboration],
  Phys.\ Rev.\ C {\bf 87}, 035204 (2013).

\bibitem{Krzemien:2014ywa}
  W.~Krzemie${\rm \acute{n}}$ {\it et al.} [WASA-at-COSY Collaboration],
  Acta Phys.\ Polon.\ B {\bf 45}, 689 (2014).
\bibitem{Krzemien:2015fsa}
  W.~Krzemie${\rm \acute{n}}$ {\it et al.} [WASA-COSY Collaboration],
  Acta Phys.\ Polon.\ B {\bf 46}, 757 (2015).

\bibitem{Skurzok:2016fuv}
  M.~Skurzok {\it et al.} [WASA-at-COSY Collaboration],
  Acta Phys.\ Polon.\ B {\bf 47}, 503 (2016).
\bibitem{Adlarson:2016dme}
  P.~Adlarson {\it et al.},
  Nucl.\ Phys.\ A {\bf 959}, 102 (2017).

\bibitem{Chrien:1988gn}
  R.~E.~Chrien {\it et al.},
  Phys.\ Rev.\ Lett.\  {\bf 60}, 2595 (1988).

\bibitem{Johnson:1993zy}
  J.~D.~Johnson {\it et al.},
  Phys.\ Rev.\ C {\bf 47}, 2571 (1993).

\bibitem{Sokol:1998ua}
  G.~A.~Sokol, T.~A.~Aibergenov, A.~V.~Kravtsov, A.~I.~L'vov, and L.~N.~Pavlyuchenko,
  Fizika B {\bf 8}, 85 (1999).

\bibitem{Budzanowski:2008fr}
  A.~Budzanowski {\it et al.} [COSY-GEM Collaboration],
  Phys.\ Rev.\ C {\bf 79}, 012201 (2009).

\bibitem{Moskal:2010ee}
  P.~Moskal and J.~Smyrski,
  Acta Phys.\ Polon.\ B {\bf 41}, 2281 (2010).

\bibitem{Pheron:2012aj}
  F.~Pheron {\it et al.},
  Phys.\ Lett.\ B {\bf 709}, 21 (2012).


\bibitem{Fujioka:2015pla}
  H.~Fujioka {\it et al.} [Super-FRS Collaboration],
  Acta Phys.\ Polon.\ B {\bf 46}, 127 (2015).


\bibitem{Rausmann:2009dn}
  T.~Rausmann {\it et al.},
  Phys.\ Rev.\ C {\bf 80}, 017001 (2009).

\bibitem{Ikeno:2017xyb}
  N.~Ikeno, H.~Nagahiro, D.~Jido and S.~Hirenzaki,
  Eur.\ Phys.\ J.\ A {\bf 53}, 194 (2017).

\bibitem{Skurzok:2018paa}
  M.~Skurzok, P.~Moskal, N.~G.~Kelkar, S.~Hirenzaki, H.~Nagahiro and N.~Ikeno,
  Phys.\ Lett.\ B {\bf 782}, 6 (2018).

\bibitem{Sick:2014yha}
  I.~Sick,
  Phys.\ Rev.\ C {\bf 90}, 064002 (2014).

\bibitem{Oset:1997it}
  E.~Oset and A.~Ramos,
  Nucl.\ Phys.\ A {\bf 635}, 99 (1998).

\bibitem{Landay:2016cjw}
  J.~Landay, M.~D\"{o}ring, C.~Fern\'{a}ndez-Ram\'{i}rez, B.~Hu and R.~Molina,
  Phys.\ Rev.\ C {\bf 95}, 015203 (2017).

\bibitem{Perez:2014jsa}
  R.~Navarro P\'erez, J.~E.~Amaro and E.~Ruiz Arriola,
  Phys.\ Lett.\ B {\bf 738}, 155 (2014).

\end{thebibliography}

\end{document}